\documentclass[preprint,12pt]{elsarticle}

\usepackage{amssymb}

\journal{Nuclear Physics B}

\begin{document}

\begin{frontmatter}


\title{Stability of black holes in Ho\v{r}ava gravity: Gravitational quasinormal modes}
\author{Nijo Varghese\corref{cor1}\fnref{label2}}
\ead{nijovarghese@cusat.ac.in}
\author{V C Kuriakose\corref{cor2}}
\ead{vck@cusat.ac.in}
\cortext[cor1]{corresponding author}
\address{Department of Physics, Cochin University of Science and Technology,
Cochin - 682 022, Kerala, India.\fnref{label3}}





\begin{abstract}
Gravitational perturbations in the geometry of asymptotically flat
black hole solution in the deformed Ho\v{r}ava-Lifshitz(HL) gravity
are considered and the associated quasinormal mode(QNM) frequencies
are evaluated using WKB method. Black holes are found to be stable
against small perturbations. We also find a significant deviation in
the behavior of QNMs in HL theory for low values of HL parameter
$\omega$, from that in the standard General Relativity(GR). QNMs are
long lived in Ho\v{r}ava theory compared with the usual
Schwarzschild black hole case.
\end{abstract}


\end{frontmatter}


\section{Introduction}
\label{intro}

One of the greatest goals in theoretical physics is to realize a
quantum version of gravity. Attempts are being made in the past to
explain gravity in the framework of quantum field theory but none
are satisfactory. Recently a renormalizable theory of gravity in 3+1
dimensions was proposed by Ho\v{r}ava\cite{horava}, inspired from
the Lifshitz theory in solid state physics, now known as
Ho\v{r}ava-Lifshits theory. The theory is a potential candidate of
quantum field theory of gravity. It assumes a Lifshitz-like
anisotropic scaling between space and time at short distances,
characterized by a dynamical critical exponent z = 3 and thus
breaking the Lorentz invariance. While in the IR limit it flows to z
= 1, retrieving the Einstein's GR. Even though the discussions on
some issues with the theory are going
on\cite{pang,coupling,inconsistency,instability}, as a new theory,
it is interesting to investigate its various aspects in parallel.
Cosmological
implications\cite{cosmo1,cosmo2,cosmo3,cosmo4,cosmo5,cosmo6} and
astrophysically viable tests of the
theory\cite{solidangle,solar,planet,solarorbit} had been studied.
Since the theory has the same Newtonian and post Newtonian
corrections as those of GR, systems of strong gravity, like black
holes, are needed to get observable deviation from the standard GR.
The black hole solutions in HL theory were studied
in\cite{lmp,ks,park,extremalbh,bhsln,topobh} and some of its
properties were investigated
in\cite{TDbh,entrobh,konoplya2,chenjing,myung,dcj,accretion}.

Stability of black holes in HL theory is the topic of this paper.
There are strong indirect evidences for the existence of black
holes, and if the black holes are found to be unstable in the new
theory it may question the reliability of the theory itself.

Gravitational perturbation and stability studies of black holes were
originally formulated by Regge and Wheeler\cite{regwhe,vish}.
Considering small perturbations in the space time outside the
horizon and assuming harmonic time dependence to this perturbation,
seek the solutions of the perturbation equation by imposing the
boundary conditions that the perturbations will vanish at the
horizon and at spatial infinity. If it admits solutions with damping
amplitudes(frequencies with negative imaginary part) then the black
hole is stable, otherwise the perturbations will grow in time and
the gravitational waves will carry away energy of the system leading
to an unstable black hole. This complex frequency with negative
imaginary part is called quasinormal modes and its existence proves
the stability of the black hole. Also features of the black holes in
the new theory would be imprinted on the QNMs and if the
gravitational detectors can observe them it can test the validity of
the theory.

The paper is organized as follows. In the following section we
briefly review asymptotically flat black hole solution in modified
HL theory and derive the radial equation for perturbation in that
black holes spacetime. In section\ref{sec2} we evaluate the QNMs.
The main conclusions are discussed in section\ref{sec4}.
\section{Gravitational perturbation around black hole in Ho\v{r}ava theory}
\label{sec2} The IR vacuum of pure HL gravity is found to be anti-de
Sitter\cite{lmp,park}. Even though HL gravity could recover GR in IR
at the action level for a particular value of the parameter
$\lambda=1$, there found a significant difference between these
black hole solutions and the usual Schwarzschild AdS. The asymptotic
fall-off of the metric for these black hole solutions is much slower
than that of usual Schwarzschild AdS black holes in GR. Meanwhile
Kehagias and Sfetsos\cite{ks} could find a black hole solution in
asymptotically flat Minkowski spacetimes applying deformation in HL
theory by adding a term proportional to the Ricci scalar of
three-geometry, $\mu^{4}R^{(3)}$ and then taking the limit
$\Lambda_{W}\rightarrow0$.  This will not alter the UV properties of
the theory but it does the IR ones leading to Minkowski vacuum
analogous to Schwarzschild spacetime in GR. The deformed action in
this particular limit is given as\cite{ks},

\begin{eqnarray}
S= \int dtd^{3}x\sqrt{g}N\Bigg\{\frac{2}{k^{2}}(K_{ij}K^{ij}-\lambda K^{2})-\frac{%
k^{2}}{2\omega ^{4}}C_{ij}C^{ij}+\frac{k^{2}\mu }{2\omega
^{2}}\epsilon
^{ijk}R_{il}^{(3)}\nabla _{j}R_{k}^{(3)l} %
\nonumber\\-\frac{k^{2}\mu ^{2}}{8}R_{ij}^{(3)}R^{(3)ij}+\frac{k^{2}\mu ^{2}}{%
8(1-3\lambda )}\frac{1-4\lambda }{4}(R^{(3)})^{2}+\mu
^{4}R^{(3)}\Bigg\} \label{eq1},
\end{eqnarray}

with second fundamental form $K_{ij}$ defined as:
\begin{equation}
\
K_{ij}=\frac{1}{2N}\left(\dot{g}_{ij}-\nabla_{i}N_{j}-\nabla_{j}N_{i}\right),
\label{eq3}
\end{equation}
and Cotton tensor $C^{ij}$,
\begin{equation}
\
C^{ij}=\epsilon^{ikl}\nabla_{k}\left(R^{(3)j}_{l}-\frac{1}{4}R^{(3)}\delta^{j}_{l}\right).
\label{eq3}
\end{equation}

Considering the metric ansatz for a static, spherically symmetric
solution,
\begin{equation}
\ ds^{2}=-N(r)^{2}dt^{2}+f(r)^{-1}dr^{2}+r^{2}d\Omega ^{2},
\label{eq2}
\end{equation}

the Lagrangian after angular integration reduces to
\begin{equation}
\ \mathcal{L}=
\frac{k^{2}\mu^{2}}{8(1-3\lambda)}\frac{N}{\sqrt{f}}\left[(2\lambda-1)\frac{(f-1)^{2}}{r^{2}}-2\lambda\frac{f-1}{r}f'+\frac{\lambda-1}{2}f'^{2}-2\omega\left(1-f-rf'\right)\right].
\label{eq3}
\end{equation}

In the IR limit $\lambda=1$ of the theory, one gets asymptotic flat
solution

\begin{equation}
\ N(r)^{2}=f(r)=1+\omega r^{2}-\sqrt{r(\omega^{2}r^{3}+4 \omega M)},
\label{coeff}
\end{equation}
with horizons at,
\begin{equation}
\ r_{\pm}=M\left(1\pm\sqrt{1-\frac{1}{2\omega M^{2}}}\right),
\label{eq4}
\end{equation}
where M is the integration constant related to ADM mass. In order
that the singularity at $r=0$ not to be naked it requires $\omega
M^{2}\geq\frac{1}{2}$. In the limit $\omega M^{2}\gg1$ this solution
recovers Schwarzschild solution in conventional General Relativity.
Metric $g_{\mu\nu}$ with coefficients given in Eq.(\ref{coeff}) is
our initial time-independent equilibrium system up on which we
consider small perturbation $h_{\mu\nu}$ with harmonic time
dependence $e^{-ikt}$. Where $k$ is the frequency. Now at late time,
we want to see whether this frequency has an imaginary part which
makes the perturbations to grow exponentially in time(unstable) or
leads to its damping(stable).

Equation governing the perturbation is the Einstein's equation for
this perturbed spacetime,
\begin{equation}
R_{\mu\nu}(g+h)=0, \label{eq5}
\end{equation}
where $R_{\mu\nu}(g+h)$ is the Ricci tensor computed from the total
metric $g_{\mu\nu}+h_{\mu\nu}$. Since the perturbation is small,
taking terms up to linear in $h_{\mu\nu}$, above equation can be
expanded as,
\begin{equation}
R_{\mu\nu}(g)+\delta R_{\mu\nu}(h)=0, \label{eq6}
\end{equation}
where $R_{\mu\nu}(g)$ is the Ricci tensor computed from the
unperturbed metric $g_{\mu\nu}$ which will vanish and the equation
becomes,

\begin{equation}
\delta R_{\mu\nu}(h)=0. \label{delr}
\end{equation}

Variation of Ricci tensor can be found from the
expression\cite{eisen}
\begin{equation}
\delta R_{\mu\nu}=-\delta
\Gamma^{\beta}_{\mu\nu;\beta}+\delta\Gamma^{\beta}_{\mu\beta;\nu},
\label{eq7}
\end{equation}
where the variation of affine connections is
\begin{equation}
\delta\Gamma^{\alpha}_{\beta\gamma}=\frac{1}{2}g^{\alpha\nu}(h_{\beta\nu;\gamma}+h_{\gamma\nu;\beta}-h_{\beta\gamma;\nu}),
\label{eqn4}
\end{equation}

Any arbitrary perturbations can be decomposed in to normal modes,
since the background we are considering is spherically symmetric.
For any given value of the angular momentum $L$, associated with
these modes, there are two classes of perturbations. Even,
$(-1)^{L}$ and odd, $(-1)^{L+1}$ parity perturbations. Here we
consider odd parity case. The canonical form of odd wave
perturbations in Regge-Wheeler gauge\cite{regwhe} is
\begin{equation}
\ h_{\mu\nu}=\begin{array}{|cccc|}
               0 & 0 & 0 & h_{0}(r) \\
               0 & 0 & 0 & h_{1}(r) \\
               0 & 0 & 0 & 0 \\
               h_{0}(r) & h_{1}(r) & 0 & 0
             \end{array}\textbf{  }e^{-ikt}(sin(\theta)\frac{\partial}{\partial\theta}P_{L}(cos(\theta)))
             . \label{odd}
\end{equation}

Substituting Eq.(\ref{odd}) in Eq.(\ref{delr}) and separating the
angular and radial parts we get the following radial equations:
\begin{equation}
\frac{ikh_{0}(r)}{f}+\frac{d}{dr}\left[fh_{1}(r)\right]=0,\textmd{
from }\delta R_{23}, \label{eqn6}
\end{equation}

\begin{eqnarray}
\frac{ik}{f(r)}\left(\frac{dh_{0}}{dr}-\frac{2h_{0}}{r}\right)+h_{1}\Bigg\{\frac{L(L+1)}{r^{2}}-\frac{k^{2}}{f}
+\frac{f}{r}\Bigg[\frac{2}{r} \qquad\qquad\qquad\qquad
\nonumber\\+\frac{2\omega(r-\frac{m+r^{3}\omega}{\sqrt{r\omega(4M+r^{3}\omega)}})}{f}\Bigg]\Bigg\}
=0, \textmd{  from  }\delta R_{13}. \label{eqn7}
\end{eqnarray}

Defining $dr_{*}=\frac{1}{f}dr$ and $Q=\frac{fh_{1}}{r}$, the above
two equations can be combined to get a second order equation by
eliminating $h_{0}(r)$:
\begin{equation}
\frac{d^{2}Q}{dr_{*}^{2}}+(k^{2}-V_{eff})Q=0. \label{eqn8}
\end{equation}
where
\begin{equation}
\ V_{eff}=f\left[\frac{L(L+1)}{r^{2}}
-6\omega\left(1-\frac{m+r^{3}\omega}{r\sqrt{r\omega(4M+r^{3}\omega)}}\right)\right]
\label{eqn9}
\end{equation}

\begin{figure}[h]
\centering
\includegraphics[width=0.6\columnwidth]{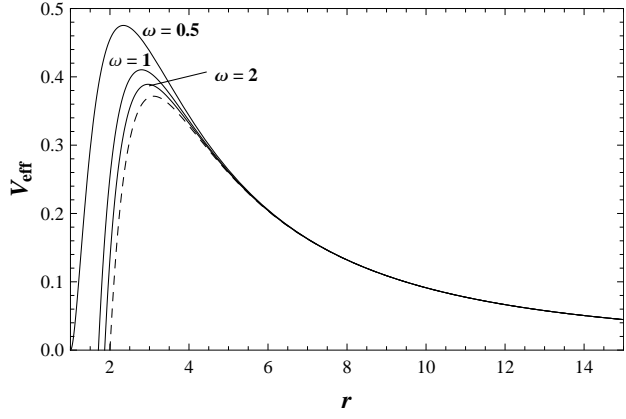}%
\caption{Effective potential for different values of $\omega$ with L=3. Dotted line represents the Schwarzschild case.} %
\label{fig1}
\end{figure}

Effective potential is plotted in Figure\ref{fig1} for $L=3$. The
height of the potential decreases with the increase of the parameter
$\omega$ and finally merges with the Schwarzschild potential as
$\omega$ tends to $\infty$.

\section{Evaluation of quasinormal modes}
\label{sec3} Third order WKB approximation method is used here to
evaluate the complex normal mode frequencies of black hole, a semi
analytic method originally developed by Schutz and
Will\cite{schutzwill} for the lowest order. Later Iyer and
Will\cite{iyerwill} carried out this approach to third WKB order and
Konoplya\cite{konoplya} to sixth order to get more accurate results.
The method is found to be accurate for low lying modes and can be
used to explore the QNM behavior of black holes efficiently, without
going to the complicated numerical methods.

WKB method gives a simple condition which will lead to discrete,
complex values for the normal mode frequencies:

\begin{equation}
\
k^{2}=[V_{0}+(-2V_{0}^{"})^{\frac{1}{2}}\Lambda]-i(n+\frac{1}{2})(-2V_{0}^{"})^{\frac{1}{2}}(1+\Omega),
\label{eqn}
\end{equation}

where $\Lambda$ and $\Omega$ are second and third order WKB
correction terms,
\begin{equation}
\Lambda =\frac{1}{(-2V_{0}^{^{\prime \prime }})^{1/2}}\left[ \frac{1}{8}%
\left( \frac{V_{0}^{(4)}}{V_{0}^{^{\prime \prime }}}\right) \left( \frac{1}{4%
}+\alpha ^{2}\right) -\frac{1}{288}\left( \frac{V_{0}^{^{\prime
\prime \prime }}}{V_{0}^{^{\prime \prime }}}\right) ^{2}\left(
7+60\alpha ^{2}\right) \right] ,  \label{a26}
\end{equation}

\bigskip
\begin{eqnarray}
\Omega &=&\frac{1}{(-2V_{0}^{^{\prime \prime
}})}\{\frac{5}{6912}\left( \frac{V_{0}^{^{\prime \prime \prime
}}}{V_{0}^{^{\prime \prime }}}\right)
^{4}\left( 77+188\alpha ^{2}\right) -\frac{1}{384}\left( \frac{%
V_{0}^{^{\prime \prime \prime }2}V_{0}^{^{(4)}}}{V_{0}^{^{\prime \prime }3}}%
\right) \left( 51+100\alpha ^{2}\right)  \nonumber \\
&&+\frac{1}{2304}\left( \frac{V_{0}^{(4)}}{V_{0}^{^{\prime \prime
}}}\right)
^{2}\left( 67+68\alpha ^{2}\right) +\frac{1}{288}\left( \frac{%
V_{0}^{^{\prime \prime \prime }}V_{0}^{^{(5)}}}{V_{0}^{^{\prime \prime }2}}%
\right) \left( 19+28\alpha ^{2}\right)  \nonumber \\
&&+\frac{1}{288}\left( \frac{V_{0}^{^{(6)}}}{V_{0}^{^{\prime \prime }}}%
\right) \left( 5+4\alpha ^{2}\right) \},  \label{a27}
\end{eqnarray}
$\alpha=n+\frac{1}{2}$ and $n$ is the mode number,
\begin{equation}
n=\{%
\begin{array}{c}
0,1,2,..................Re(E)\succ 0 \\
-1,-2,-3,...............Im(E)\prec 0.%
\end{array}
\label{eqn11}
\end{equation}

$V_{0}^{(n)}$ denotes the n$^{th}$ derivative of $V$ evaluated at
$r_{0}$, the value of $r$ at which $V$ attains maximum.
Table\ref{table1} and Table\ref{table2} list the QNMs evaluated.

\begin{table}[h]
\scriptsize
\begin{tabular}{ccccc}
\hline\hline $\omega $ & L=2 & L=3 & L=4 & L=5 \\ \hline 0.5 & \ \
0.44322 - 0.05945 i \  &  \ 0.68305 - 0.06558 i  & \ \ 0.91050 -
0.06764 i  &  \ 1.13252\ - 0.06863 i \  \\
1 & 0.40001 - 0.07875 i & 0.63226 - 0.08339 i & 0.84942 - 0.08506 i
&
1.06027 - 0.08586 i \\
1.5 & 0.38949 - 0.08221 i & 0.61995 - 0.08694 i & 0.83446 - 0.08857
i &
1.04250 - 0.08933 i \\
2 & 0.38488 - 0.08397 i & 0.61434 - 0.08853 i & 0.82762 - 0.09011 i
&
1.03435 - 0.09087 i \\
3 & 0.38065 - 0.08573 i & 0.60904 - 0.09001 i & 0.82113 - 0.09156 i
&
1.02662 - 0.09229 i \\
4 & 0.37866 - 0.08661 i & 0.60650 - 0.09072 i & 0.81801 - 0.09224 i
&
1.02290 - 0.09297 i \\
6 & 0.37676 - 0.08749 i & 0.60403 - 0.09140 i & 0.81497 - 0.09290 i
&
1.01927 - 0.09362 i \\
8 & 0.37583 - 0.08792 i & 0.60281 - 0.09174 i & 0.81348 - 0.09323 i
&
1.01748 - 0.09394 i \\
10 & 0.37528 - 0.08819 i & 0.60209 - 0.09194 i & 0.81259 - 0.09342 i
&
1.01642 - 0.09413 i \\
$\infty $ & 0.37316 - 0.08922 i & 0.59927 - 0.09273 i & 0.80910 -
0.09417 i & 1.02568 - 0.09067 i \\ \hline\hline
\end{tabular}
\caption{Fundamental$(n=0)$ QNMs of gravitational perturbations for
various values of $\omega$ and L. $\omega=\infty $ represents the
Schwarzschild limit}  \label{table1}
\end{table}
 \normalsize

\begin{table}[h]
\scriptsize
\begin{tabular}{cccccc}
\hline\hline $n$ & $\omega $=1 & $\omega $=2 & $\omega $=3 & $\omega
$=4 & $\omega =\infty $ \\ \hline 0 & \ 0.63226 - 0.08339 i\  & \ \
0.61434 - 0.08853 i & \ \ 0.60904 -
0.09001 i\  & \ \ 0.60650 - 0.09072 i & \ \ 0.59927 - 0.09273 i \  \\
1 & 0.62029 - 0.25228 i & 0.59996 - 0.26827 i & 0.59381 - 0.27290 i
&
0.59085 - 0.27511 i & 0.58235 - 0.28141 i \\
2 & 0.59924 - 0.42583 i & 0.57504 - 0.45373 i & 0.56747 - 0.46182 i
&
0.56380 - 0.46568 i & 0.55320 - 0.47668 i \\
3 & 0.57207 - 0.60378 i & 0.54310 - 0.64417 i & 0.53372 - 0.65589 i
&
0.52913 - 0.66149 i & 0.51575 - 0.67743 i \\
4 & 0.54009 - 0.78467 i & 0.50538 - 0.83773 i & 0.49372 - 0.85315 i
& 0.48797 - 0.86053 i & 0.47107 - 0.88154 i \\ \hline\hline
\end{tabular}
\caption{Higher mode frequencies for different values of $\omega$
with L=3} \label{table2}
\end{table}

 \normalsize

\begin{figure}[h]
\centering
\includegraphics[width=0.45\columnwidth]{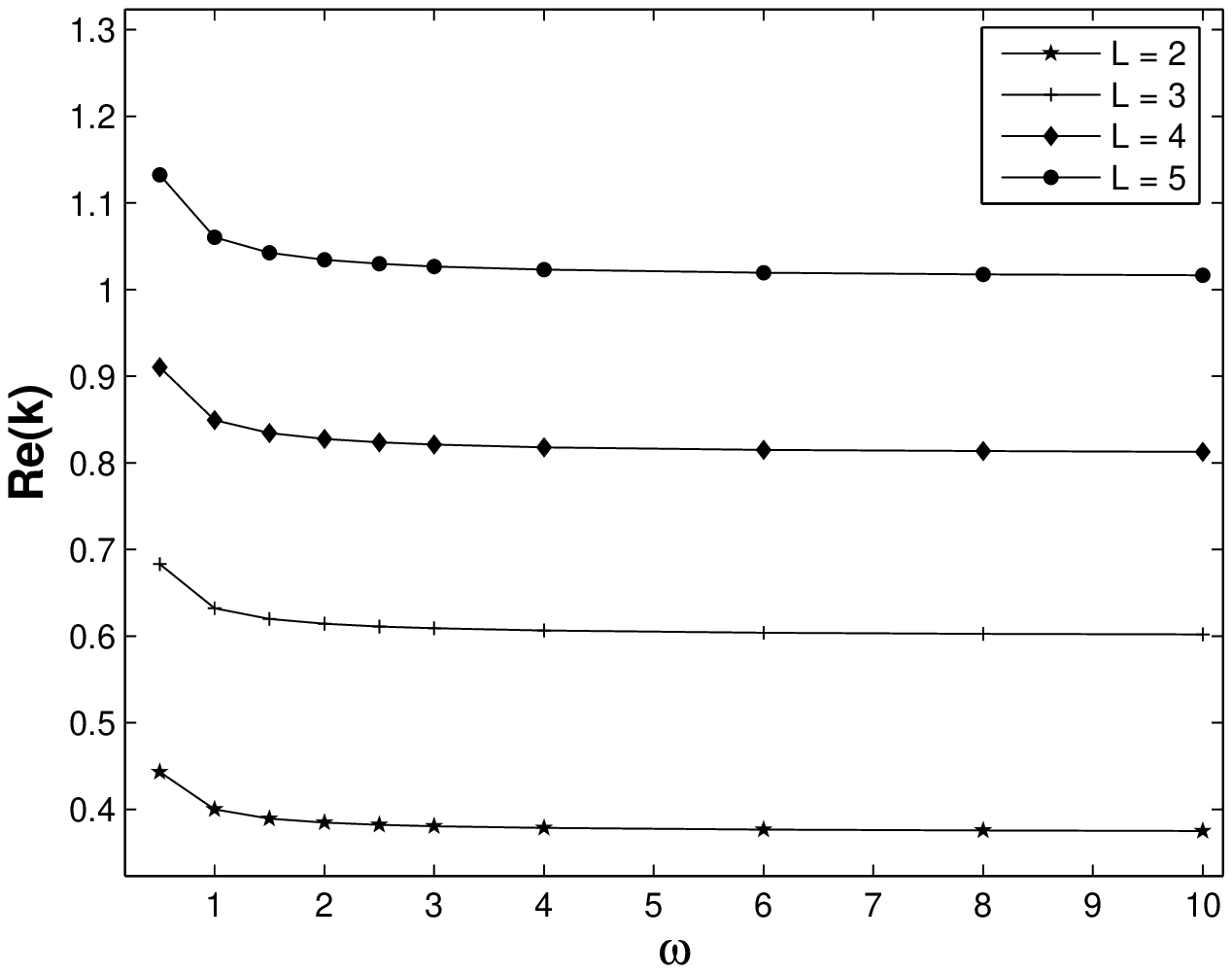}%
\hspace{0.1in}%
\includegraphics[width=0.45\columnwidth]{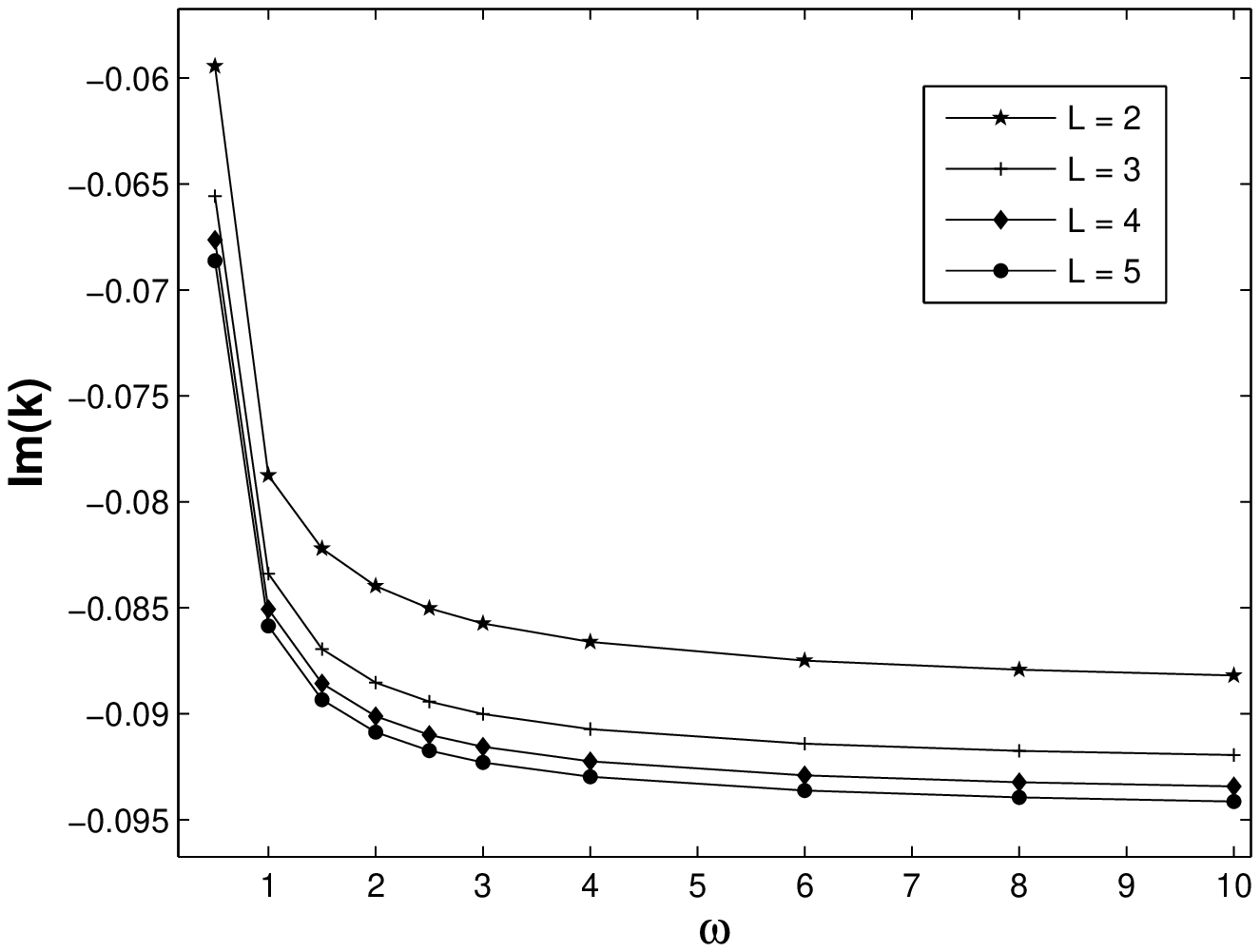}
\caption{$Re(k)$ and $Im(k)$ as a function of $w$ for $n=0$ and
L=2,3,4,5.
} %
\label{fig2}
\end{figure}

The variation of real and imaginary parts of QNMs with the parameter
$\omega$ are plotted in Figure\ref{fig2}. The oscillation frequency,
$Re(\omega)$ decreases with the increase of $w$ while the damping
time $|Im(\omega)|$, increases with the increase of $w$. For values
near $w=0.5$ there is a sudden decrease in damping time. Finally
comparing with the Schwarzschild black hole, oscillation frequencies
of QNMs in HL gravity is larger and has a lower damping time.

\section{Conclusion}
\label{sec4} We have investigated the evolution of gravitational
perturbations in black hole spacetime in HL theory and associated
QNMs are evaluated. The negative imaginary part of QNMs show that
the perturbation will decay in time and hence we can conclude that
black holes are stable against small perturbations in the spacetime.
The present calculations show that the gravitational QNMs are long
lived in Ho\v{r}ava theory compared with the Schwarzschild black
hole.

\section*{Acknowledgements}
NV wishes to thank UGC, New Delhi for financial support under RFSMS
scheme. VCK is thankful to CSIR, New Delhi for financial support
through a Research Scheme and wishes to acknowledge Associateship of
IUCAA, Pune, India.





\bibliographystyle{elsarticle-num}






\end{document}